# The height of chitinous ridges alone produces the entire structural colour palette

*Hemant Kumar Raut[1,†], Qifeng Ruan[1,†], Cédric Finet[2,3], Vinodkumar Saranathan[2,3,4], Joel K.W. Yang[1], Javier G. Fernandez[1]\**

1 Engineering and Product Development, Singapore University of Technology and Design, Singapore
2 Department of Biological Sciences, National University of Singapore, Singapore
3 Division of Science, Yale-NUS College, National University of Singapore, Singapore
4 Division of Sciences, School of Interwoven Arts and Sciences, Krea University, Sricity, India

***Corresponding author**: javier.fernandez@sutd.edu.sg (JGF)
†Equal contribution
H.K.R. now at École polytechnique fédérale de Lausanne, Switzerland
Q.R. now at Harbin Institute of Technology, China

## Abstract

The colourful wings of butterflies result from the interaction between light and the intricate chitinous nanostructures on butterflies' scales. This study demonstrates that just by reproducing the chitinous ridges present in butterfly scales (i.e., without any other secondary structure), the entire colour palette is achieved. This result was achieved using a new methodology based on the controlled reproduction of parts of the biological structure of complex chitinous systems using their native chemistry, enabling the isolation of different features' contributions. Here we isolate the contribution of the ridges and their variations as producing and modulating colour hue. The results suggest that complicated butterfly scales may be non-ideal solutions for producing colour when multifunctionality is not considered.

**Main**

Structural colour in arthropod cuticles, particularly on butterfly wings, is one of the most ubiquitous and striking examples of ostentation in nature, and the nanopatterns producing these colours have been extensively studied[1]. It is generally assumed that the primary function of these colour-producing nanostructures in butterfly wing scales is to produce colour; therefore, their extreme complexity is required for such a task. However, this rather obvious assumption may be erroneous. Butterfly scales are a multipurpose convolution of features resulting from constraints and needs that may or may not be related to the many factors involved in the generation of colour[2]. Therefore, identifying the particular contributions of each feature of the scales to determine their functionality is central to understanding the physiological and evolutionary aspects of butterflies.

In bioengineering, the motivation to understand the biology of butterflies is derived from the goal to artificially reproduce structural colour in chitinous objects[3, 4]. Knowing that butterfly scales offer the solution to this, the aim of this study was to determine the basic principles behind this solution, hidden within the multipurpose 'noise' of the colour-producing structures.

Over a decade ago, we demonstrated that chitinous polymers retain their ability to crystallise and form nanostructures after extraction from the cuticle. This ability can be explored to form topographies of a few hundred nanometres[5, 6]. To explore the characteristics of chitinous structural colour, we built upon those results, achieving the control and quality necessary to make use of the optical properties of the material. This enabled the reproduction of the structural chitinous ridges without the rest of the scales' topography, effectively isolating their contribution to the production of colour.

We used two-photon lithography to produce ridge structures[7] with a horizontal spacing of 3.4 µm to match the interridge distance across several genera[8, 9] (**Fig. 1a**). We then mapped the achievable colour produced at the fixed interridge spacing by altering the vertical dimensions from 150 nm to 2.4 µm. We kept the interridge distance constant because of the consistency of the horizontal interridge distances across butterfly species and colours[10]. We focused on the variation of the ridge height[11] inspired by recent evidence supporting the idea that the alteration of the vertical dimension (i.e., the thickness of the lower lamina) is a previously unconsidered evolutionary strategy to explore different colourations in butterflies[12-15].

The structures produced in the synthetic photocurable resin were transferred to a silicone elastomer through soft lithography and transformed into chitinous structures through a process that involved casting chitosan (i.e., highly deacetylated chitin), which had been extracted from shrimp shells and dispersed in a weak acetic acid solution (**Fig. 1b**). The controlled evaporation of the solvent resulted in freestanding chitinous films containing a reproduction of the original ridge-like nanotopography with a height scaled by a factor of 0.73 by the vertical shrinking of chitosan during crystallisation[16] (**Fig. 1c**).

These experiments demonstrated that the otherwise transparent chitosan films (**Supplementary Fig. 1a**) started to show colour when the ridges were under 400 nm in height, and they covered the whole spectrum in the next 1.4 µm. Then, as the height continued to increase, the colour sequence started repeating (**Figs. 2a, 2b**, **Supplementary Fig. 1b**). Despite the strong and lineal correlation between the ridge height and the colour hue, when these results were reproduced for similar ranges of interridge distances and ridge widths, those parameters showed negligible influence on the resulting hue compared to the height (**Fig. 3a**).

To obtain intuitive insight into the origin of the peaks and dips in the transmission spectra, we simulated the structures and their effects on a plane wave normally incident on the chitinous gratings (**Fig. 3b, Supplementary Fig. 2**). The system can be understood as a thin-film interference reflector, such as those commonly occurring in butterfly scales [17, 18], in which two waves propagate through chitinous polymer and air. This interpretation results in simulated transmission spectra remarkably close to those measured in the physical samples. When the phase difference between the two paths equals an even (odd) integer of $\pi$, the constructive (destructive) interference of the two waves gives rise to large (small) transmittance[7]. The phase difference between the two paths increases with the increment in ridge height, resulting in the redshift in the peaks and dips, which we have reported as a correlation between the height of the grating and the colour hue. The slight discrepancy between the measured and simulated spectra could be ascribed to the deviation in the shape of the ridges from the ideal cuboid and the variation in the refractive index from 1.56 at some wavelengths[19]. It is worth noting that in the theoretical reproduction of the colour palette, the supporting chitinous layer is not involved, which is in agreement with the independence of the colour hue with the thickness of the supporting layer in the fabricated structures. This is because the phase difference between the two waves is not altered when both waves pass through the supporting layer of constant thickness before travelling through the chitinous polymer/air-structured area. This could also be observed on the flat spectra obtained in the unstructured and low structures (i.e., < 0.3 µm) surfaces (both theoretical and fabricated).

This is the first demonstration of the achievement of any colour using only the heights of chitinous ridges and without any of the structures typically associated with the production of colour, such as lower lamina, honeycombs or lumen multilayers and lattices[20] (**Fig. 1a**). Therefore, the independent contribution of ridge heights to the overall colour of butterfly wing scales should be carefully integrated into future studies. Our findings support the idea that the diversity and complexity of the colour-producing structures in butterfly wings—and by extension, the cuticles of other arthropods—cannot be solely explained as an efficient strategy for producing different colours[2], as this feat can be achieved by simpler and more efficient design variations. At the same time, other functionalities of the cuticle, such as further optical properties (iridescence, angle independency, light polarisation, etc.)[21], interactions with water[22], aerodynamics[23], structural requirements or even the winding evolutionary path to these structures[24], need to be considered when studying and replicating biological structures that produce colour.

**Materials and Methods**

**Materials**

Chitosan (medium molecular weight, high degree of deacetylation; Sigma Aldrich), 3-(trimethoxysilyl) propyl methacrylate (TMSPMA), acetic acid and NaOH were used as received. Polydimethylsiloxane (PDMS) mould with the trench patterns were prepared using a SYLGARD 184 silicon elastomer kit.

**Fabrication of patterned chitosan surface**

The two-photon polymerization lithography (TPL)-based 3D printing of the colour producing trench arrays was performed as described elsewhere.[7] The trench arrays were casted with PDMS and cured to transfer the negative replica to a PDMS soft mould. The latter was then used to cast chitosan and fabricate trench array containing chitosan films. The chitosan-based resist was prepared by dissolving 3% w/v chitosan power in 1% v/v acetic acid.

To measure the spectra, the chitosan trenches were also fabricated on glass. In case of the latter, high fidelity pattern transfer to glass substrate became possible by enhancing the adhesion of the chitosan resist to the glass substrate. This was performed by a surface functionalization of glass using TMSPMA that enables chitosan to adhere strongly to glass.[25]. For soft lithography, a few drops of the chitosan resist were dispensed on glass substrate and pressed with a PDMS mould comprising the topographical pattern. The soft imprinting was conducted until the chitosan film was dried after which the PDMS mould was peeled off. Subsequently, the substrate with chitosan trenches was submerged in NaOH 4% (w/v) for 10 min to neutralize the protonated amino groups and avoid further dissolution.[6] Finally, the substrate was washed in deionized water to remove any remaining NaOH and dried at 37°C.

### Optical Measurements

The optical microscopy images and transmittance spectra were measured using a Nikon Eclipse LV100ND optical microscope equipped with a Nikon DS-Ri2 camera and a CRAIC 508 PV microspectrophotometer. Two halogen lamps (LV-HL 50 W) were used to illuminate the samples in the reflection/transmission mode. A 5×, NA = 0.15 objective was employed to measure transmission spectra. The spectra were normalized to the transmittance spectrum of the non-patterned area of the chitosan film.

### Characterization

Scanning electron microscopy (SEM) was performed using a JEOL-JSM-7600F and JEOL JSM 6010LV SEM system (Jeol Ltd., Tokyo) with an accelerating voltage of 5 kV and 15-20kV, respectively. Platinum sputter-coated butterfly wing scale samples were milled using a gallium ion beam on a FEI Versa 3D with the following settings: beam voltage - 8kV, beam current – 12 pA at a 52° tilt and imaged on the same equipment using a beam voltage of 5kV, and beam current of 13 pA. Light microscope images of individual butterfly wing scales were recorded using the 100X lens of a uSight-2000-Ni microspectrophotometer (Technospex Pte. Ltd., Singapore) fitted with a Touptek U3CMOS-05 camera.

### Simulation

The transmission spectra of the chitinous gratings were calculated with a finite-difference time-domain software (Lumerical Solution). Chitinous polymer was modeled with a refractive index of 1.56. A plane wave was normally incident the chitinous gratings with periodic boundary conditions. A field and power monitor was placed above the gratings. Considering the numerical aperture (0.15) of the experimentally used objective lens, a near-to-far-field projection was conducted to integrate the transmitted light within an 8.6° cone. The calculated spectra were then normalized to a reference transmission spectrum of a semi-infinite chitinous substrate without the ridges.

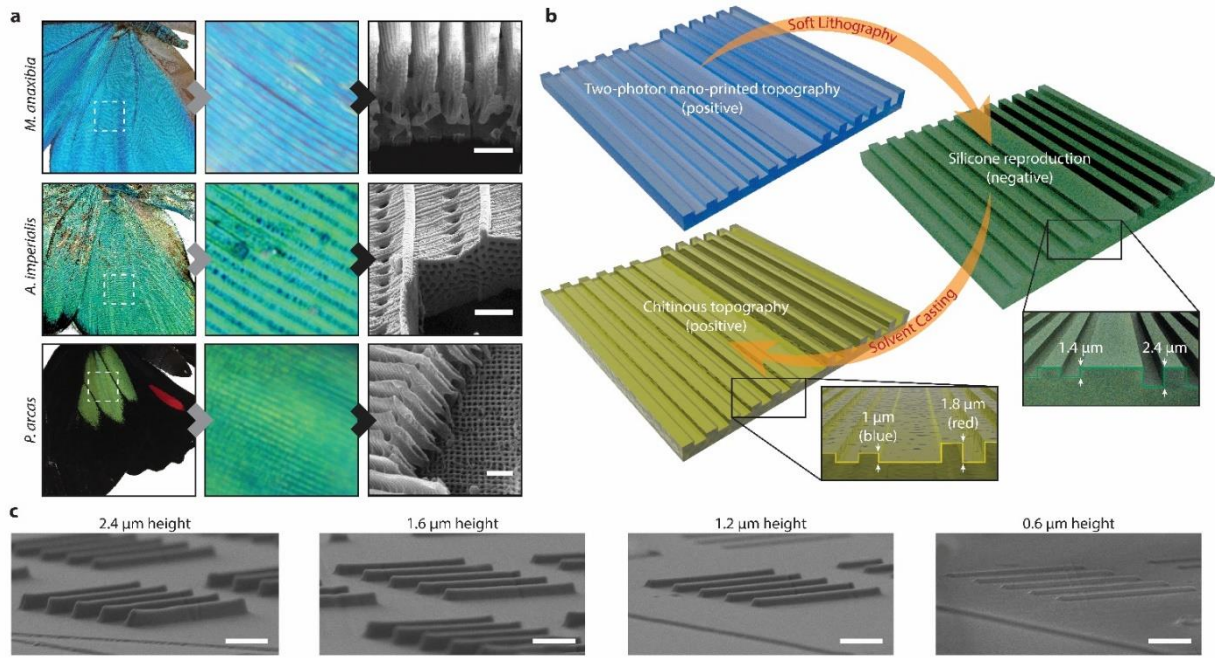

**Figure 1. Fabrication of chitinous nano-ridges. a)** Colour-producing structures in butterfly wings are characterised by high levels of complexity and diversity. These panels show representative examples from different Lepidoptera: *Morpho anaxibia* (Nymphalidae, top row), *Arcas imperialis* (Lycaenidae, middle row) and *Parides arcas* (Papilionidae, bottom row). Photos of the butterfly (left), light micrographs captured at 100x magnifications (middle) and scanning electron microscope (SEM) images of cross-sections made with a focused ion beam (FIB) or manual fracture (right) are shown; the SEM scale bars are 1 µm. **b)** Diagram of the fabrication process and characteristics of the biomimetic chitinous ridges. The structures were initially fabricated by two-photon additive manufacturing, which enabled accurate height control. A negative copy of the structures was produced in an elastomeric polymer, which was used to cast a chitosan solution in liquid crystal form. **c)** SEM images of 0.8-µm-wide chitinous ridges of different heights; the bars are 4 µm.

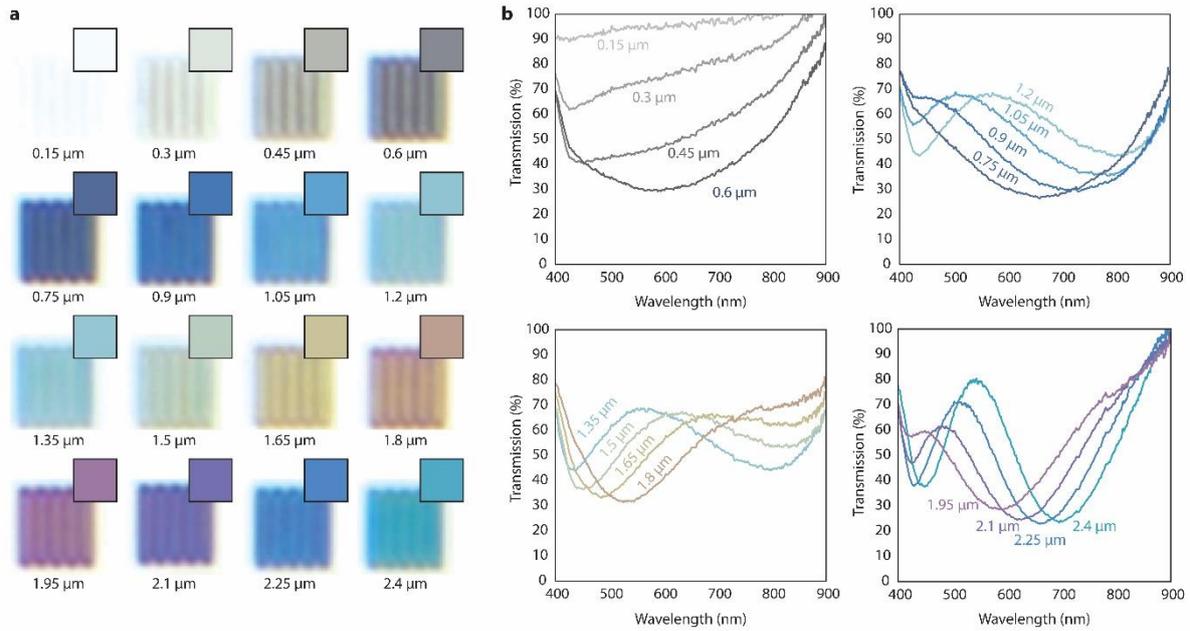

**Figure 2. Colour produced by the chitinous ridges**. **a)** The whole palette of colours produced by changing the height of the chitinous ridges from 150 nm to 2.4 µm. The images show 'pixels' of 18 x 18 µm containing five ridges of similar height taken with an optical microscope. The top-right inset in each pixel shows the averaged colour observable at the macro scale. **b)** Transmission spectra produced by ridges with heights ranging from 150 nm to 2.4 µm.

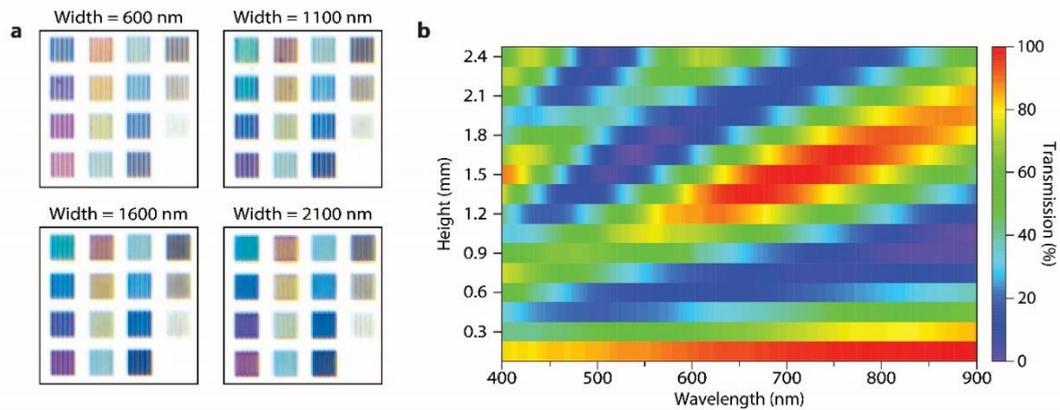

**Figure 3. Influence of other geometrical factors and simulated results**. **a)** Changing the width of the ridges from 0.6 µm (top-left) to 2.1 µm (bottom-right) at constant height did not significantly influence colour. However, changing the height of the ridges (from 0.15 µm to 2.4 µm in each of the images, similar to Fig. 2a) resulted in large, linear and continuous changes in the colour hue. **b)** Simulated results for the structures in Fig. 1 and Fig. 2 (i.e., pitch = 3.4 µm, width = 0.8 µm, height 0.15–2.4 µm) normalised to a flat (i.e., unstructured) chitinous film. The simulation closely matched the observations, where low structures (< 0.3 µm) showed an almost flat transmission spectra, which started to show sharp minima (dark blue-purple in the graph) at short wavelengths, and they were displaced to longer wavelengths as the height of the ridges increased.

**Acknowledgements**

The authors would like to thank John You En Chan for his assistance in designing the moulds, Vaishnavi Sundararajan and Gianluca Grenci (MBI) for access and help with SEM, and the Pennycook group (MSE) for the use of FIB-SEM. We also want to thank Kwi Shan Seah for the photo of *P. arcas* and Michael Greeff and the ETH Zürich Entomologische Sammlung for access to specimens, loans, and photography. This work was supported by the National Research Foundation of Singapore (grant number: CRP20-2017-0004).

# The height of chitinous ridges alone produces the entire structural colour palette

Hemant Kumar Raut†, Qifeng Ruan†, Cédric Finet, Vinodkumar Saranathan, Joel K.W. Yang, Javier G. Fernandez*

*Corresponding author: javier.fernandez@sutd.edu.sg
†Equal contribution

This material includes:

**Supplementary Fig.1** Photo of the structured chitosan film and mosaic figures made with ridges of varying heights
**Supplementary Fig.2** Example of simulated spectra.

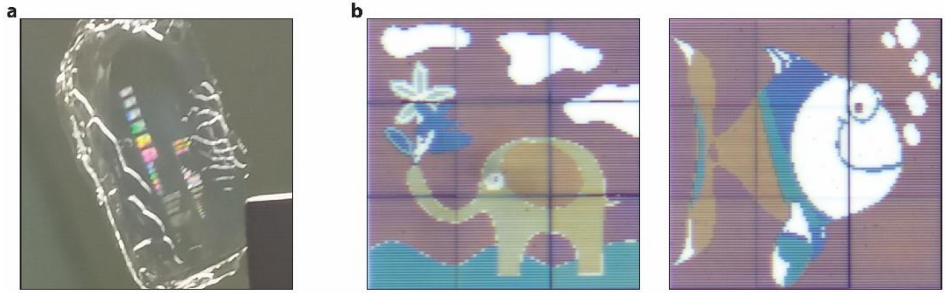

**Supplementary Fig.1|Coloured chitinous ridges. a**, Chitinous film with patches of ridges showing colour at the macroscale. The film is approximately 8×13 mm **b**, Combination of ridges of different heights to form images of 300×300 µm.

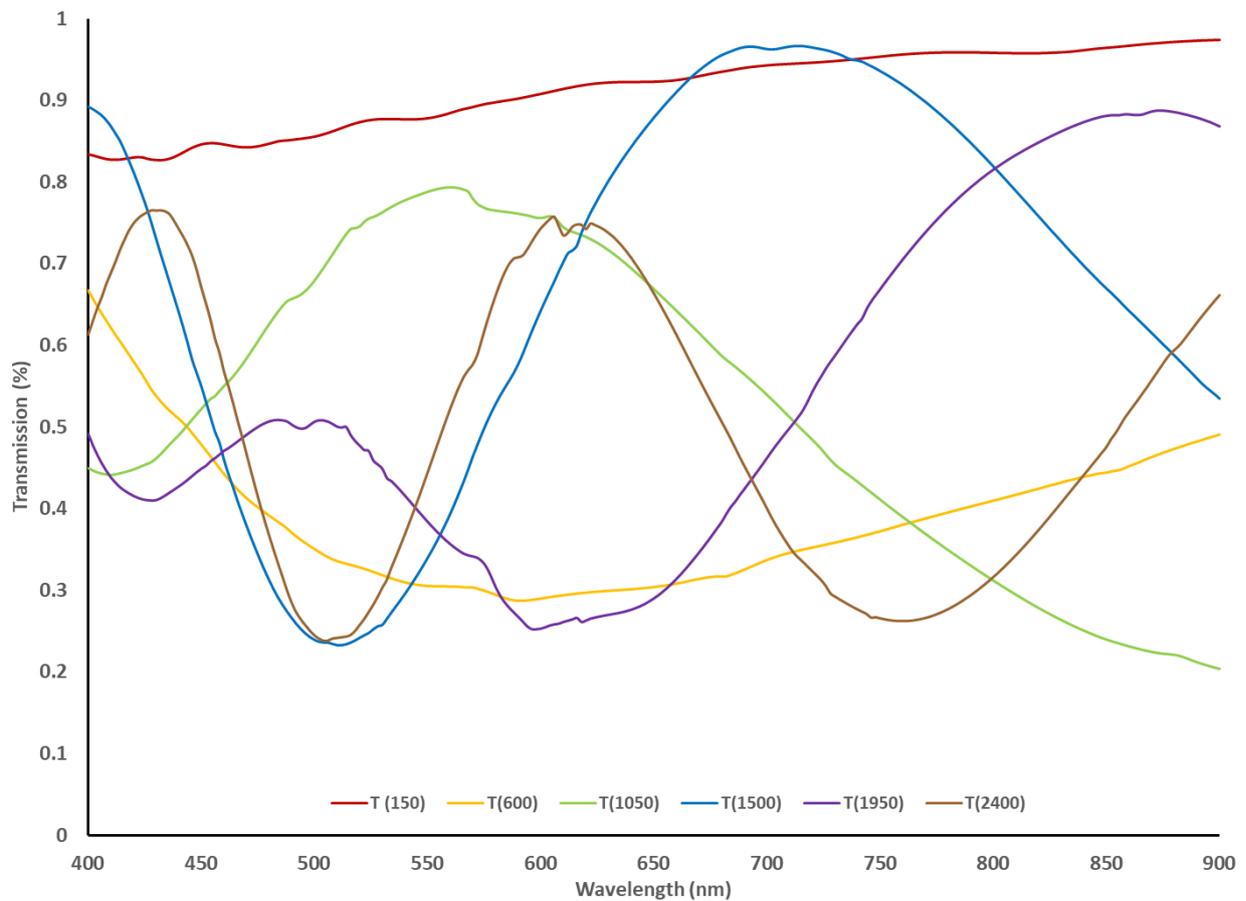

**Supplementary Fig.2| Examples of simulated spectra.** Example of a few simulated spectra produced by chitinous structures with a pitch = 3.4 um, width = 0.8 um, heights of 0.15 to 2.4 um, and a refractive index of ~ 1.56. Examples are extracted from Fig. 3b of the main article, where additional intermediate heights are included.